\pdfoutput=1
\documentclass{aa}
\usepackage{graphicx}
\usepackage{amsmath,amssymb,amsfonts}
\usepackage{txfonts}
\usepackage{booktabs}
\usepackage{xcolor}
\usepackage{hyperref}
\hypersetup{hidelinks}
\usepackage{orcidlink}
\usepackage{tikz}
\usetikzlibrary{arrows.meta, positioning, fit, backgrounds, calc}

\begin{document}

   \title{Partial-field inferability of galaxy overdensity structure}
   \subtitle{An information-geometric account of incomplete sky coverage}

   \author{Marko Imbri\v{s}ak\inst{1}\orcidlink{0000-0002-2773-8617}%
   \thanks{\emph{marko.imbrisak@gmail.com}}
   \and Kre\v{s}imir Tisani\'c\inst{1}\orcidlink{0000-0001-6382-4937}}

   \institute{
     Independent Researcher, Zagreb, Croatia
   }

   \date{}

  \abstract
   {
     Structural statistics of the galaxy distribution are routinely
     measured on only partially observed fields. Yet the amount of
     structure that survives incomplete coverage, and which aspects are
     lost first, is rarely quantified independently of the particular
     statistic being reported.
   }
   {
     We cast this question as inference on a fixed structure vector
     and evaluate recovery with a Fisher-inspired reference metric. From the
     Voronoi tessellation of a galaxy field, we construct the
     three-component overdensity structure vector $\Theta=(\Delta_\Omega,
     f_{\rm HD}, L_{\rm HD})$, comprising a global density contrast, a
     high-density fraction, and the connected extent of the
     high-density set. We then ask with what probability a subset of
     the field recovers $\Theta$ within a prescribed information
     tolerance.
   }
   {
     We assign local densities to galaxies through their Voronoi cell
     areas and propagate connectivity on the dual Delaunay graph. Two
     mechanisms generate subsets: spatially correlated random-walk
     coverage and spatially uncorrelated independent node loss. Recovery
     is
     assessed with a regularised Mahalanobis distance
     $d_g^2=\Delta\Theta^{\rm T} g_\Theta \Delta\Theta$, using a single
     reference metric estimated once from a stated deviation ensemble
     and held fixed across coverage levels, mechanisms and estimators.
     This yields an inferability $H_\varepsilon$ and a survivability
     $S_\varepsilon$ that can be compared at matched coverage. We apply the
     analysis to a
     $z$-sliced COSMOS field containing $94\,043$ galaxies.
   }
   {
     The three components form a strict hierarchy. At every incomplete
     coverage we sample, the connected extent $L_{\rm HD}$ is recovered
     several times less accurately than the scalar components, by a
     margin that widens as coverage improves, and it is the last to
     converge. The two subset mechanisms trade off in opposite
     directions: uniform loss preserves the scalar amplitudes more
     accurately, whereas walk coverage better preserves morphology.
     Inferability therefore depends not only on how much of the field is
     observed, but also on how it is observed. A joint criterion
     combining the three components can order the two mechanisms at
     matched coverage, but that single-number ordering is tolerance
     dependent and can reverse, whereas the component-level contrast is
     independent of the metric and robust. We further show that the
     degree-based inverse-probability weight, although natural for
     random-walk sampling, is inconsistent for set-valued coverage and
     leaves a bias that does not vanish even at complete coverage.
   }
   {
     Partial-field structure statistics therefore degrade
     anisotropically in information space: scalar amplitudes
     survive sampling that destroys connectivity, and connectivity
     survives sampling that biases amplitudes. A single coverage
     fraction is thus insufficient to characterise what a partially
     observed field can support.
   }


   \maketitle
   \nolinenumbers

   \section{Introduction}
   \label{sec:intro}

Statistics of the galaxy distribution are almost never measured over a
complete field. Masks, bright-star holes, chip gaps, variable depth,
and finite survey boundaries ensure that the galaxies contributing to
a measurement form only a subset of the underlying population
\citep{colless2001, swanson2008}. The
usual response is to report a coverage or completeness fraction
alongside the result. That number, however, describes the data rather
than the inference: it states how much of the field was observed, not
how much of its structure remains recoverable under that pattern of
observation.

The distinction matters because structural statistics do not degrade
uniformly. A mean density averages over its members and so tolerates
many losses; a connected extent depends on which surviving members
remain adjacent, and removals that scarcely shift any average can break
it. Two subsets of identical size may therefore support very different
conclusions, even though a single coverage fraction assigns them the
same nominal completeness.

We therefore treat partial coverage as an inference problem. Let a
\emph{structure vector} $\Theta$ (a fixed list of structure
statistics, reported together and treated as one object) encode the
structural content of interest. For a specified mechanism that
generates observed subsets, we ask how often an estimator based only
on the subset falls within a prescribed tolerance of the full-field
value. The tolerance is measured in a metric that places all
components of $\Theta$ on a common scale. The resulting probability,
viewed as a function of both coverage and sampling mechanism, defines
the \emph{inferability} of $\Theta$.

This formulation is adapted from distributed information storage on
networks, where the same construction appears with reversed roles.
There, one asks how much of a stored secret an adversary can recover
through partial access. That probability is its \emph{hackability}:
the chance that partial access suffices to reconstruct what was
stored. One asks in the same way how much survives partial
destruction, its \emph{survivability}, the probability that the stored
object remains recoverable after part of the network is removed. The
two are then combined into a single figure of merit
\citep{zlatic2026}. In the
astrophysical analogue, the observer takes the role of the adversary,
incomplete sky coverage corresponds to partial access, and asking what
a partial field supports becomes equivalent to asking what a partial
network reveals. The preceding reconstruction study identified this
connection as an open direction \citep{imbrisak_paperII}; here we
make it quantitative.

We construct the structure vector from the Voronoi tessellation of
the galaxy field. Tessellation-based density estimation has a long
history in this setting \citep{icke1987, vandeweygaert1994,
schaap2000}: the inverse area of a galaxy's Voronoi cell provides an
adaptive, parameter-free estimate of local density, while the dual
Delaunay triangulation supplies the natural adjacency through which
that density becomes connected structure. From this construction we
combine three quantities: a global density contrast, the fraction of
galaxies classified as high-density, and the connected extent of the
largest high-density component. Together they span a progression from
pure amplitude to pure morphology.

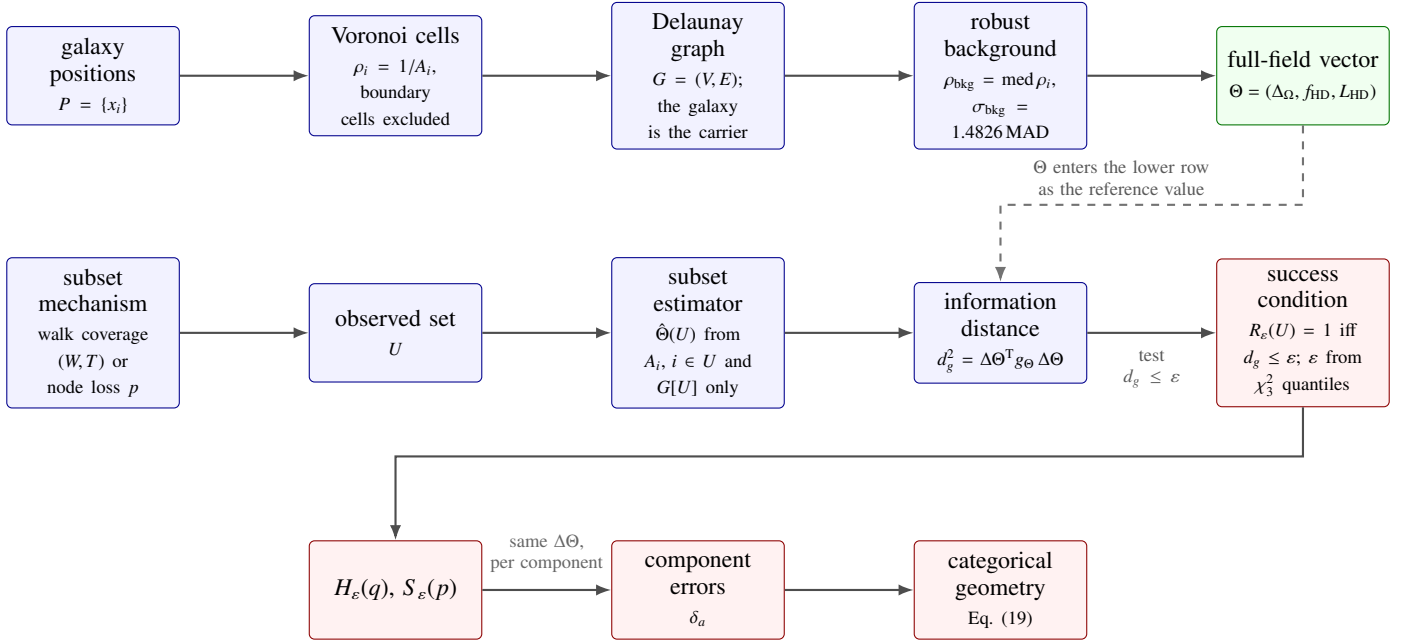
\begin{figure*}
\centering
\begin{tikzpicture}[
   font=\small,
   box/.style   = {draw=blue!55!black, fill=blue!5, rounded corners=2pt,
                   align=center, inner sep=4pt, text width=20mm,
                   minimum height=13mm},
   outbox/.style= {draw=red!55!black, fill=red!5, rounded corners=2pt,
                   align=center, inner sep=4pt, text width=20mm,
                   minimum height=13mm},
   refbox/.style= {draw=green!45!black, fill=green!6, rounded corners=2pt,
                   align=center, inner sep=4pt, text width=20mm,
                   minimum height=13mm},
   ar/.style    = {-{Latex[length=2mm]}, draw=black!70, thick},
   dashar/.style  = {-{Latex[length=2mm]}, draw=black!55, thick, dashed}
]
\def\cA{0}\def\cB{4.0}\def\cC{8.0}\def\cD{12.0}\def\cE{16.0}
\def\rI{0}\def\rII{-3.4}\def\rIII{-6.8}

\node[box]    (pts) at (\cA,\rI) {galaxy positions\\[1pt]
                                  \scriptsize $P=\{x_i\}$};
\node[box]    (vor) at (\cB,\rI) {Voronoi cells\\[1pt]
                                  \scriptsize $\rho_i=1/A_i$, boundary
                                  cells excluded};
\node[box]    (del) at (\cC,\rI) {Delaunay graph\\[1pt]
                                  \scriptsize $G=(V,E)$; the galaxy is
                                  the carrier};
\node[box]    (bkg) at (\cD,\rI) {robust background\\[1pt]
\scriptsize $\rho_{\rm bkg}=\mathrm{med}\,\rho_i$,
                                  $\sigma_{\rm bkg}=1.4826\,$MAD};
\node[refbox] (thq) at (\cE,\rI) {full-field vector\\[1pt]
                                  \scriptsize $\Theta=(\Delta_\Omega,
                                  f_{\rm HD}, L_{\rm HD})$};
\foreach \a/\b in {pts/vor, vor/del, del/bkg, bkg/thq} \draw[ar] (\a) -- (\b);

\node[box]    (mec) at (\cA,\rII) {subset mechanism\\[1pt]
                                   \scriptsize walk coverage $(W,T)$
                                   or node loss $p$};
\node[box]    (obs) at (\cB,\rII) {observed set\\[1pt] \scriptsize $U$};
\node[box]    (est) at (\cC,\rII) {subset estimator\\[1pt]
                                   \scriptsize $\hat\Theta(U)$ from
                                   $A_i,\, i\in U$ and $G[U]$ only};
\node[box]    (dst) at (\cD,\rII) {information distance\\[1pt]
                                   \scriptsize $d_g^2=\Delta\Theta^{\rm T}
                                   g_\Theta\,\Delta\Theta$};
\node[outbox] (suc) at (\cE,\rII) {success condition\\[1pt]
                                   \scriptsize $R_\varepsilon(U)=1$ iff
                                   $d_g\le\varepsilon$; $\varepsilon$ from
                                   $\chi^2_3$ quantiles};
\foreach \a/\b in {mec/obs, obs/est, est/dst} \draw[ar] (\a) -- (\b);
\draw[ar] (dst) -- (suc)
   node[midway, below, yshift=-1mm, font=\scriptsize, text=black!65,
        text width=15mm, align=center]
   {test\\ $d_g\le\varepsilon$};

\node[outbox] (cur) at (\cB,\rIII) {$H_\varepsilon(q)$, $S_\varepsilon(p)$};
\node[outbox] (cmp) at (\cC,\rIII) {component errors\\[1pt]
                                    \scriptsize $\delta_a$};
\node[outbox] (cat) at (\cD,\rIII) {categorical geometry\\[1pt]
                                    \scriptsize Eq.~(\ref{eq:cat})};
\draw[ar] (cur) -- (cmp)
   node[midway, above, yshift=1mm, font=\scriptsize, text=black!60,
        text width=16mm, align=center]
   {same $\Delta\Theta$,\\ per component};
\draw[ar] (cmp) -- (cat);

\draw[dashar] (thq.south) -- ++(0,-1.05) -| (dst.north)
   node[pos=0.30, above, font=\scriptsize, text=black!65,
        text width=34mm, align=center]
   {$\Theta$ enters the lower row as the reference value};
\draw[ar] (suc.south) -- ++(0,-0.65) -| (cur.north);
\end{tikzpicture}
\caption{Flow of the partial-field analysis. The upper row builds the
reference: the Voronoi tessellation gives each galaxy a cell area and
hence a local density, the dual Delaunay graph supplies the adjacency
along which that density is connected, and a robust background fixes
the high-density threshold, yielding the full-field structure vector $\Theta$.
The lower row is the experiment: a subset mechanism produces an
observed set $U$, the estimator forms $\hat\Theta(U)$ from
$U$-information alone (Sect.~\ref{sec:contract}), and the deviation is
measured in the reference metric to give the success indicator
$R_\varepsilon$. The dashed arrow marks the comparison with the
full-field structure vector
$\Theta$, which enters as the reference value in $d_g$; the component
errors $\delta_a$ are that same comparison read component by
component. The metric $g_\Theta$ is estimated separately, from the
stated calibration ensemble, and then held fixed
(Sect.~\ref{sec:metric}). Averaging
$R_\varepsilon$ gives the inferability and survivability curves, while
the component errors $\delta_a$ carry the mechanism comparison without
reference to $g_\Theta$ or $\varepsilon$.}
\label{fig:flow}
\end{figure*}

The paper is organised as follows. Section~\ref{sec:data} describes the
COSMOS field and the working
patch. Section~\ref{sec:method}, whose procedure is summarised in the
schematic of Fig.~\ref{fig:flow}, sets out the density construction, the
structure vector, the subset
estimators and the information restriction they obey, the reference
metric and success criterion, the two subset mechanisms, and the
categorical geometry that links inferability and survivability.
Section~\ref{sec:results} presents the measured curves, the component
hierarchy, the mechanism dependence, and a consistency check that
rules out the naive sampling correction. Sections~\ref{sec:discussion}
and~\ref{sec:conclusions} discuss the implications and summarise.

   \section{Data}
   \label{sec:data}

\subsection{COSMOS photometric-redshift slice}
\label{sec:data-field}

We use sky positions from a COSMOS-field galaxy catalogue
\citep{smolcic2007}, restricted to a photometric-redshift slice
$0.120 \le z_{\rm phot} \le 0.320$. The slice contains $94\,043$
galaxies spanning right ascension
$09^{\rm h}57^{\rm m}39^{\rm s}$ to $10^{\rm h}03^{\rm m}18^{\rm s}$
and declination $+01\degr30\arcmin$ to $+02\degr55\arcmin$, an area of
about $2\,{\rm deg}^2$.

The redshift cut serves a specific purpose. Because the density
estimate in Sect.~\ref{sec:dens} is two-dimensional, projecting the
full catalogue depth would superpose physically unrelated structures
within a single tessellation and thereby dilute the connected
morphology we aim to track. The chosen slice is narrow enough to limit
that mixing, yet broad enough to retain a useful surface density, so
the projected tessellation remains a meaningful proxy for structure
at a common epoch. We do not correct for redshift errors within the
slice. Instead, the analysis treats the resulting point set as the
field of interest and asks what partial coverage can recover from it.

\subsection{Working patch}
\label{sec:data-patch}

The full field is used to establish the geometric regime
(Sect.~\ref{sec:res-field}). The inferability analysis itself is run
on a contiguous patch of $N=2000$ galaxies, taken as the $N$ nearest
neighbours of the field centre, about $0\fdg23$ on a side and
$0.05\,{\rm deg}^2$ in area.
The patch size balances geometric content against computational cost.
It must be large enough for the high-density set to develop
non-trivial connected structure; otherwise the morphological component
of the structure vector would be statistically empty. At the same
time, it must remain small enough for the Monte Carlo analysis, which
requires $10^3$--$10^4$ independent subset realisations, to remain
inexpensive. At $N=2000$ a full analysis takes approximately three
minutes for the
$4.9\times10^4$ subset realisations reported here.

The patch is nested inside the field, and both are built with
identical geometric conventions, so results at patch scale and field
scale refer to the same construction. The preceding reconstruction
study \citep{imbrisak_paperII} used a much smaller cut of the same
catalogue ($N=119$); the relation between the two is discussed in
Sect.~\ref{sec:res-field}.

   \section{Method}
   \label{sec:method}

\subsection{Density field and its carrier}
\label{sec:dens}

Let $P=\{x_i\}_{i=1}^{N}$ be the set of galaxy positions in the plane
of the sky. The Voronoi tessellation assigns to each galaxy the region
of the plane closer to it than to any other galaxy,
\begin{equation}
   V_i = \{ x : \lVert x-x_i \rVert \le \lVert x-x_j \rVert
   \ \ \forall j \},
\end{equation}
and the local surface density carried by galaxy $i$ is the inverse of
the area of that cell,
\begin{equation}
   \rho_i = 1/A_i, \qquad A_i = \lvert V_i \rvert .
   \label{eq:rho}
\end{equation}

The carrier convention must be stated explicitly because the dual
tessellation permits a second, inequivalent assignment. Density in
Eq.~(\ref{eq:rho}) belongs to the \emph{galaxy}, not to the Voronoi
vertex: a cell corner is a geometric feature of the tessellation with
no mass associated to it. The adjacency along which density is
compared must therefore be the adjacency of galaxies, which is the
Delaunay triangulation $G=(V,E)$ dual to the tessellation, and not the
Voronoi vertex-adjacency graph. Both graphs are available from the
same construction and both were used as reconstruction test-beds in
\citet{imbrisak_paperII}, but only the Delaunay graph carries the
density field, and all structure statistics below are computed on it.

Cells at the edge of the point set are unbounded. We reconstruct them
against a survey window taken as the bounding box of $P$, which
renders every area finite, but a window-limited area is a property of
the window rather than of the field, so boundary galaxies are excluded
from the density carrier. They remain in the triangulation, where they
contribute adjacency, but they do not enter $\Theta$. We write $V^{\rm
int}$ for the interior set and $n=\lvert V^{\rm int} \rvert$.

The background level against which overdensity is defined is estimated
internally, as
\begin{equation}
   \rho_{\rm bkg} = \mathrm{med}_{\,i \in V^{\rm int}}\, \rho_i ,
   \qquad
   \sigma_{\rm bkg} = 1.4826\,
   \mathrm{med}_{\,i \in V^{\rm int}}\, \lvert \rho_i - \rho_{\rm bkg}
   \rvert .
   \label{eq:bkg}
\end{equation}
The scale is the median absolute deviation rescaled by $1.4826$, the
consistency factor that makes it agree with the standard deviation for
normally distributed data \citep{hampel1974, rousseeuw1993}. A robust
estimator is required here rather than preferred.
The distribution of $1/A_i$ is strongly right-skewed: on
our patch, the mean exceeds the median by a factor of $1.48$. A
mean-based background would therefore be pulled toward the densest
cells and would no longer represent the baseline against which those
same cells are meant to stand out. We return to the consequences of this
choice in
Sect.~\ref{sec:res-field}.

\subsection{The overdensity structure vector}
\label{sec:theta}

For an observed set $U \subseteq V^{\rm int}$, we define three
complementary quantities. The first is a median-referenced global
density contrast,
\begin{equation}
   \Delta_\Omega(U) = \frac{\rho_\Omega(U)}{\rho_{\rm bkg}} - 1,
   \qquad
   \rho_\Omega(U) = \frac{\sum_{i\in U} w_i}{\sum_{i \in U} w_i A_i} ,
   \label{eq:delta}
\end{equation}
with weights $w_i$ discussed in Sect.~\ref{sec:ipw} and $w_i \equiv 1$
in the unweighted case. Equation~(\ref{eq:delta}) is an area-weighted
density and is deliberately \emph{not} the arithmetic mean of the
$\rho_i$; the two differ by the skewness of the cell-area
distribution, and conflating them is the most common way to
misestimate a tessellation-based density.

The second is the fraction of observed galaxies flagged as
high-density,
\begin{equation}
   f_{\rm HD}(U) = \frac{\sum_{i \in U} w_i h_i}{\sum_{i \in U} w_i},
   \qquad
   h_i =
   \begin{cases}
      1, & \rho_i > \rho_{\rm bkg} + k_{\rm HD}\,\sigma_{\rm bkg},\\
      0, & \text{otherwise},
   \end{cases}
   \label{eq:fhd}
\end{equation}
where $k_{\rm HD}$ sets the threshold in units of the background
scatter.

The third quantity captures morphology. Let $G[U \cap \mathcal{H}]$ be the
Delaunay graph restricted to the observed high-density galaxies
$\mathcal{H}=\{i : h_i=1\}$, and let $\mathcal{C}$ be its connected
components. Then
\begin{equation}
   L_{\rm HD}(U) = \frac{\max_{C \in \mathcal{C}}
   \sum_{i \in C} w_i}{\sum_{i \in U \cap \mathcal{H}} w_i}
   \label{eq:lhd}
\end{equation}
is the share of the high-density population carried by its largest
connected component, a measure of whether the
high-density set forms one coherent structure or many disjoint
fragments, with the
convention $L_{\rm HD}(U)=0$ when $U \cap \mathcal{H} = \varnothing$.

Together, these quantities define the structure vector
\begin{equation}
   \Theta(U) = \bigl(\Delta_\Omega(U),\, f_{\rm HD}(U),\,
   L_{\rm HD}(U)\bigr) \in \mathbb{R}^3 ,
   \label{eq:product}
\end{equation}
and we write $\Theta \equiv \Theta(V^{\rm int})$ for the full-field
value. The three components are chosen to span a range: $\Delta_\Omega$
is a pure amplitude and involves no adjacency at all; $f_{\rm HD}$ is
a counting statistic that involves the threshold but still no
adjacency; $L_{\rm HD}$ is a pure morphology and is defined entirely
through adjacency. Any difference in how they respond to partial
coverage is therefore attributable to that progression.

\subsection{Subset estimators and the information restriction}
\label{sec:contract}

Equations~(\ref{eq:delta})--(\ref{eq:lhd}) depend on $U$ alone. The
estimator evaluated on an observed subset
may use: the cell areas $A_i$ for $i \in U$; the Delaunay adjacency of
$G$ (Sect.~\ref{sec:dens}) restricted to $U$; and the background
constants
$(\rho_{\rm bkg},\sigma_{\rm bkg})$, which are treated as externally
supplied calibration. It may not use $A_j$ for any unobserved $j$, nor
any component of the full-field $\Theta$, nor the identity of the
unobserved set.

This restriction is what gives the resulting probabilities
their inferential meaning.
An estimator with access to the full field would trivially recover
$\Theta$ from any subset, and a study of its success rate would
measure nothing. The restriction also has a subtle consequence for the
morphological
component: because the Delaunay adjacency is restricted to $U$, a
high-density component that is connected in the full field can appear
as several fragments to an observer who is missing the galaxies that
bridge them. This is not an artefact to be corrected but precisely the
effect $L_{\rm HD}$ is designed to detect.

There is one deliberate exception. The background of Eq.~(\ref{eq:bkg}) is
estimated on the full interior set and held fixed across all subsets.
Estimating it internally on each subset would confound the question we
are asking --- how well a fixed structural statement can be recovered
--- with a separate question about calibration stability under
sampling. Treating $(\rho_{\rm bkg},\sigma_{\rm bkg})$ as external is
the cleaner experiment, and corresponds in practice to a background
calibrated from a wider or deeper reference field.

\subsection{Reference metric and information distance}
\label{sec:metric}

Because the components of $\Theta$ have different dimensions and
scales, comparing $\hat\Theta(U)$ with $\Theta$ requires a common
metric. We use a quadratic form
\begin{equation}
   d_g^2(U) = \Delta\Theta^{\rm T}\, g_\Theta \,\Delta\Theta ,
   \qquad \Delta\Theta = \hat\Theta(U)-\Theta ,
   \label{eq:dg}
\end{equation}
with
\begin{equation}
   g_\Theta = \bigl(\Sigma_\Theta + \lambda R\bigr)^{+} ,
   \qquad
   \Sigma_\Theta = \mathbb{E}\!\left[
   \Delta\Theta\,\Delta\Theta^{\rm T} \right] ,
   \label{eq:g}
\end{equation}
where $^{+}$ denotes the Moore--Penrose pseudoinverse, $R = \mathrm{diag}
(\Sigma_\Theta)$ is a ridge target and $\lambda=10^{-3}$ a
regularisation constant that keeps $g_\Theta$ well conditioned when
one component is nearly degenerate. Here $\Sigma_\Theta$ is the
second-moment matrix of the deviations $\Delta\Theta$ about the
full-field value, pooled across every coverage level of the
calibration ensemble (the sense of ``pooled'' made precise below), and
is deliberately uncentred:
$\mathbb{E}[\Delta\Theta\,\Delta\Theta^{\rm T}] =
\operatorname{Cov}(\Delta\Theta) +
\mathbb{E}[\Delta\Theta]\,\mathbb{E}[\Delta\Theta]^{\rm T}$. The
systematic part is retained because the displacement of a subset
estimator from $\Theta$ is exactly what we wish to penalise, and it is
not a small correction: for walk subsets at $90\%$ coverage the mean
deviation reaches $0.85$, $0.91$ and $0.58$ of the corresponding root
mean square, so the bias supplies most of the matrix. We therefore
call $g_\Theta$ a regularised empirical precision-like reference
metric, built by analogy with the Fisher metric of a local Gaussian
model \citep{rao1945, amari2000} rather than being one: were
$\Delta\Theta \sim \mathcal{N}(0,\Sigma_\Theta)$, $d_g^2$ would be the
Mahalanobis distance, a Gaussian idealisation we use in
Sect.~\ref{sec:success} only to fix the nominal tolerance scale.

The expectation in Eq.~(\ref{eq:g}) is taken over deviations pooled
across the entire coverage grid, not at a single coverage. This choice
is deliberate and consequential. A metric estimated at one coverage
inherits that coverage's scatter, so the resulting distance measures
deviations relative to a level of sampling rather than in absolute
terms, and the numerical value of $d_g$ ceases to be comparable across
the curve. Pooling makes $\Sigma_\Theta$ a fixed property of the patch and the
stated calibration ensemble, so that $d_g$ has one meaning throughout
and the $\chi^2_3$ quantiles provide a common nominal tolerance
scale. For the same reason we estimate
$\Sigma_\Theta$ once, from the unweighted ensemble, and use that single
reference metric for every estimator we compare. A metric re-estimated
from each estimator's own deviations would rescale the ruler along with
the quantity being measured, and distances obtained under different
estimators would not be comparable. Pooling thus defines a single
reference metric for the patch and the stated calibration ensemble,
held fixed across coverage levels, subset mechanisms and estimators,
so that every distance we report uses the same ruler.

Two properties of that ruler must be stated, because both bear on the
comparisons of Sect.~\ref{sec:results}. First, the composition of the
reference grid sets the numerical scale of $d_g$: we use the coverage
grid of Sect.~\ref{sec:mech} throughout, and coarsening it to half as
many levels rescales every distance by a nearly uniform factor of
order $15\%$, so $\varepsilon$ is to be read relative to a stated
ensemble. Second, the calibration ensemble is generated by the walk
mechanism alone. The metric is therefore shared but not neutral
between the two mechanisms of Sect.~\ref{sec:mech}: its correlation
structure is inherited from walk deviations, and relative to a
diagonal approximation the full metric inflates walk distances by
about $15\%$ but loss distances by about $50\%$ (median over
realisations of the per-realisation ratio). We return to the consequence in
Sect.~\ref{sec:res-mech}.

\subsection{Success criterion}
\label{sec:success}

An observation is judged successful when the recovered vector lies
within a tolerance,
\begin{equation}
   R_\varepsilon(U) =
   \begin{cases}
      1, & d_g(U) \le \varepsilon,\\
      0, & \text{otherwise}.
   \end{cases}
   \label{eq:R}
\end{equation}
The tolerance $\varepsilon$ is fixed on the $\chi^2_3$ scale that
$d_g^2$ would follow if the deviations were Gaussian with covariance
$\Sigma_\Theta$ (Sect.~\ref{sec:metric}): in that idealised case
$d_g^2 \sim \chi^2_3$, so a radius that encloses a chosen probability
mass of the reference is the square root of the corresponding
$\chi^2_3$ quantile,
\begin{equation}
   \varepsilon_q = \sqrt{F^{-1}_{\chi^2_3}(q)} ,
   \label{eq:eps}
\end{equation}
where $F^{-1}_{\chi^2_3}$ is the inverse cumulative distribution
function (the quantile function) of $\chi^2_3$. This is the only place
the distribution enters. We use the four quartile levels
$q=0.25,0.50,0.75,0.90$, that is $\varepsilon = 1.10, 1.54, 2.03,
2.50$, taking the median value $\varepsilon=1.54$ as the primary
criterion; these four values are the tolerance columns of
Tables~\ref{tab:H} and~\ref{tab:S}, and $\varepsilon=1.54$ sets the
operating point of Figs.~\ref{fig:curves} and~\ref{fig:karate}.
The quantiles supply a common nominal scale for the
tolerance family rather than exact confidence radii: the reference
matrix is estimated from a regularised, coverage-pooled and uncentred
deviation ensemble rather than from a single Gaussian population, and
$f_{\rm HD}$ and $L_{\rm HD}$ are bounded, so the $\chi^2_3$ law is a
calibration convention and not a sampling distribution. We report a
family of thresholds because $\varepsilon$ specifies the amount of
information loss an application is willing to tolerate; it is not an
intrinsic property of the data. A conclusion that survives
only one selected value of $\varepsilon$ is therefore not a robust
conclusion about the field.

By construction $d_g(V^{\rm int})=0$ and hence
$R_\varepsilon(V^{\rm int})=1$: an observer who sees the whole
interior recovers the full-field structure vector exactly. This identity is a
consistency requirement on the estimator, and we use it as such in
Sect.~\ref{sec:res-ipw}.

\subsection{Subset mechanisms}
\label{sec:mech}

We generate observed subsets with two mechanisms that retain a
specified fraction of the field but differ sharply in spatial
character.

\emph{Walk coverage.} $W$ independent random walkers of length $T$ are
released from uniformly chosen starting galaxies and step to a
uniformly chosen Delaunay neighbour at each of $T$ steps; $U$ is the
set of galaxies visited at least once. This is the sampling geometry of
the reconstruction problem of
\citet{imbrisak_paperII}, where the observable is built from walk
co-visitation (the number of
times the walker is seen to step from one galaxy to a given
neighbour, so that the observable retains the pair rather than only
the galaxy), and it produces spatially clustered subsets. At high
coverage the subgraph of $G$ induced on the visited set is connected, a single
component at $90\%$ coverage, but it fragments as coverage falls, to an
average of about eight components
at $38\%$ coverage, the largest holding some $59\%$ of the visited
galaxies. We therefore describe the
mechanism as spatially correlated rather than strictly contiguous. We
use $T=16$ throughout and
vary $W$ over the fifteen values listed in Table~\ref{tab:H}, chosen to
sample coverage densely through the range where recovery changes most
rapidly.

\emph{Independent node loss.} Each interior galaxy is retained
independently with probability $1-p$, so $U = V^{\rm int} \setminus F$
with $F$ a uniformly random subset of expected size $p\,n$. It is the
standard model of spatially uncorrelated random
incompleteness, with equal retention probability for every carrier.
The retention probability is uniform; the resulting subset is not,
since the underlying field is not.

The two mechanisms are compared at matched \emph{node coverage},
defined as $\lvert U \cap V^{\rm int}\rvert / n$ (the fraction of
density carriers observed, not a fraction of sky area), so that any
difference between them is attributable to the spatial character of the
subset rather
than to its size.

\subsection{Inferability and survivability}
\label{sec:HS}

For each mechanism, we define the probability of satisfying the
recovery criterion,
\begin{align}
   H_\varepsilon(q) &= \Pr\bigl[ R_\varepsilon(U)=1 \bigr],
   \quad U \sim \text{walk coverage at level } q ,
   \label{eq:H}\\
   S_\varepsilon(p) &= \Pr\bigl[ R_\varepsilon(V^{\rm int}\setminus F)
   =1 \bigr], \quad F \sim \text{loss at rate } p .
   \label{eq:S}
\end{align}
Following the terminology of \citet{zlatic2026} we call $H$ an
inferability (the analogue of hackability: how much an observer with
partial access can recover) and $S$ a survivability (how much
withstands partial destruction). These probabilities are not
complementary. They are probabilities of the same event under two
different subset distributions, and the identity $H = 1-S$ holds only
if the two mechanisms are related by complementation, which they are
not: the complement of a walk-visited set is not a uniformly random
set. Section~\ref{sec:res-mech} shows the difference is large.

Both are estimated by Monte Carlo with $M$ realisations per grid
point. Because a grid point at which every realisation fails
would otherwise enter the categorical geometry of
Sect.~\ref{sec:cat} as an exact zero, we quote throughout the
Jeffreys-smoothed estimator $(k+\tfrac12)/(M+1)$ rather than the raw
frequency $k/M$. This is the posterior mean of the success probability
under the Jeffreys prior $\mathrm{Beta}(\tfrac12,\tfrac12)$ for a
binomial proportion \citep{jeffreys1946, brown2001}, and it keeps
every estimate strictly inside $(0,1)$.

\subsection{Categorical Fisher geometry}
\label{sec:cat}

The pair $(H,S)$ together with a mixing weight $\alpha$ specifying how
much an application cares about inference versus persistence defines a
hierarchical four-outcome categorical model,
\begin{equation}
   p_1 = \alpha S, \quad p_2 = \alpha(1-S), \quad
   p_3 = (1-\alpha)H, \quad p_4 = (1-\alpha)(1-H),
   \label{eq:cat4}
\end{equation}
in which one first selects the persistence branch with probability
$\alpha$ or the inference branch with probability $1-\alpha$, and then
observes success or failure within the selected branch. The Fisher
information metric of this model is
\begin{equation}
   g_{ab} = \sum_{r=1}^{4}
   \frac{(\partial_a p_r)(\partial_b p_r)}{p_r} ,
   \label{eq:gab}
\end{equation}
and substituting Eq.~(\ref{eq:cat4}) gives the three diagonal entries
\begin{align}
   g_{\alpha\alpha} &= \frac{S}{\alpha} + \frac{1-S}{\alpha}
      + \frac{H}{1-\alpha} + \frac{1-H}{1-\alpha}
      = \frac{1}{\alpha} + \frac{1}{1-\alpha} ,
      \label{eq:gaa}\\
   g_{SS} &= \frac{\alpha}{S} + \frac{\alpha}{1-S} ,
      \label{eq:gss}\\
   g_{HH} &= \frac{1-\alpha}{H} + \frac{1-\alpha}{1-H} .
      \label{eq:ghh}
\end{align}
The cross terms cancel identically; for instance
$g_{\alpha S} = 1-1 = 0$, the two outcomes of the selected branch
contributing with opposite sign. What remains is the diagonal Fisher
metric
\begin{equation}
   \mathrm{d}s^2 = \frac{\mathrm{d}\alpha^2}{\alpha(1-\alpha)}
   + \frac{\alpha\,\mathrm{d}S^2}{S(1-S)}
   + \frac{(1-\alpha)\,\mathrm{d}H^2}{H(1-H)} .
   \label{eq:cat}
\end{equation}
This is the information-geometric completion of the scalar figures of
merit that summarise a position on the $(H,S)$ plane. Which summary is
appropriate depends on what partial access means. Where access is
adversarial, as in the distributed-storage setting of
\citet{zlatic2026}, high inferability is a liability,
\begin{equation}
   F_{\rm sec} = \alpha S + (1-\alpha)(1-H),
   \label{eq:fsec}
\end{equation}
whereas for an observer of a partially covered field high inferability
is the goal,
\begin{equation}
   F_{\rm astro} = \alpha S + (1-\alpha)\,H .
   \label{eq:fastro}
\end{equation}
The geometry of Eq.~(\ref{eq:cat}) is the same in both readings: only
the sign with which $H$ enters the utility changes, and
Appendix~\ref{app:sec} demonstrates this on a small network with a
security-flavoured structure vector. Equation~(\ref{eq:cat}) records
how
sharply distinguishable neighbouring positions on that plane are. The
metric diverges as $H$ or $S$ approaches $0$ or $1$: near the
boundaries, neighbouring probabilities are sharply distinguishable in
Fisher distance. Such regimes nevertheless carry little practical
trade-off, because recovery there is nearly deterministic.

\subsection{Sampling correction}
\label{sec:ipw}

Walk coverage does not sample galaxies uniformly: a random walk on an
unweighted graph visits galaxy $i$ with stationary probability
$\pi_i \propto \deg(i)$ \citep{newman2018}, where the degree
$\deg(i)$ is the number of Delaunay neighbours of that galaxy, so a
subset built from walk visits over-represents high-degree
galaxies. The textbook remedy is
inverse-probability weighting \citep{horvitz1952}, $w_i = 1/\pi_i$,
applied in Eqs.~(\ref{eq:delta})--(\ref{eq:lhd}).

This is the wrong correction here, and the reason is instructive. The
weight $1/\pi_i$ corrects a sample drawn \emph{with} the stationary
frequency, one in
which a galaxy visited ten times contributes ten times. Our observed set
is not that object: it is the
set of galaxies visited \emph{at least once}, in which a galaxy
contributes once however often it was traversed. The inclusion
probability of the set-valued observable --- the chance
that a given galaxy is observed at all --- is
$\pi^{\rm incl}_i = \Pr[i \in U]$, which tends to $1$ for every $i$ as
coverage grows, whereas $1/\pi_i$ tends
to a non-trivial constant. Applying the frequency weight to a
set-valued sample therefore introduces a bias that does not vanish
with complete coverage, and it breaks the consistency identity of
Sect.~\ref{sec:success}. We verify this in Sect.~\ref{sec:res-ipw} and
adopt the unweighted
estimator as the reference. The first-order inclusion probability $w_i =
1/\pi^{\rm incl}_i$,
which must be estimated from the same Monte Carlo ensemble, is the
appropriate design weight (the survey-statistics correction
that restores an unbiased average when units enter the sample with
unequal probability) for the scalar set-valued target quantities
$\Delta_\Omega$ and $f_{\rm HD}$, and we leave it to future work.
Correcting the topology-dependent component
$L_{\rm HD}$ is a separate problem, and one that no per-galaxy weight
can solve: reweighting the observed
galaxies cannot reconstruct an unobserved bridge, so it would require
joint inclusion probabilities
or an explicit model for the missing connectivity.

We also report the effective sample size $N_{\rm eff} = (\sum_i
w_i)^2/\sum_i w_i^2$, which for weighted estimators quantifies how
much of the nominal subset size survives the weighting.

\subsection{Relation to the reconstruction pipeline}
\label{sec:recon}

The analysis above uses the true Delaunay graph, constructed directly
from the catalogue positions and therefore known exactly. As a result,
its errors arise from partial observation alone, independently of any
graph-reconstruction step. A
companion question, which we do not treat here, is what happens when
the adjacency itself is inferred rather than known.

That setting is the subject of \citet{imbrisak_paperII}, where a
frame-balanced Levenberg--Marquardt scheme (fbLM) recovers the
adjacency of a graph from the co-visitation matrix of random walks
upon it, reaching a Matthews correlation coefficient of $0.988$ on the
COSMOS Delaunay graph in the finite-walk regime. The relevant conclusion
of that work for present purposes is that its
residual errors are almost entirely confined to edges the walk never
traversed, and on the Delaunay graph exactly so, so that
reconstruction fidelity there is governed by walk coverage rather than
by the estimator. Since walk
coverage is also the mechanism controlling
$H_\varepsilon$ here, the two effects are coupled, and separating them
requires running the present analysis on a reconstructed adjacency
$\hat{G}$ with subset realisations matched to those on $G$. We
outline that experiment in Sect.~\ref{sec:discussion}.

   \section{Results}
   \label{sec:results}

\subsection{The field and its overdensity structure vector}
\label{sec:res-field}

The Voronoi tessellation of the full slice yields $94\,020$ bounded
cells out of $94\,043$ galaxies, so only $0.02\%$ of the field is
boundary-limited, and the Delaunay triangulation has $282\,103$ edges.
The boundary fraction is a strong function of field size, since it
scales as the ratio of perimeter to area: it is $10.9\%$ for the
$N=119$ cut used in \citet{imbrisak_paperII}, $3.4\%$ at $N=1000$,
$1.9\%$ at $N=2000$, $1.1\%$ at $N=5000$, and negligible on the full
field. The wider field is thus not merely more data but a
qualitatively cleaner geometric regime, in which the exclusion of
boundary carriers (Sect.~\ref{sec:dens}) is a marginal correction
rather than a structural feature of the analysis.

\begin{figure}
   \centering
   \includegraphics[width=\hsize]{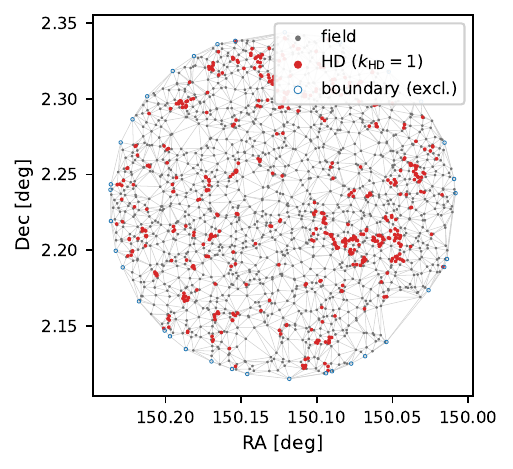}
   \caption{The $N=2000$ working patch. Grey lines are Delaunay edges,
   grey points are field galaxies, red points are high-density
   galaxies at $k_{\rm HD}=1$, and open blue circles mark boundary
   galaxies excluded from the density carrier. The high-density set is
   spatially fragmented, which is what makes its connected extent
   $L_{\rm HD}$ a sensitive probe of partial coverage.}
   \label{fig:patch}
\end{figure}

On the $N=2000$ patch (Fig.~\ref{fig:patch}) there are $1963$ interior
carriers and $5960$ Delaunay edges. The background estimate of
Eq.~(\ref{eq:bkg}) gives
$\rho_{\rm bkg}=5.65\times10^{4}\,{\rm deg}^{-2}$ and
$\sigma_{\rm bkg}=3.73\times10^{4}\,{\rm deg}^{-2}$, against an
arithmetic mean of $8.35\times10^{4}\,{\rm deg}^{-2}$; the
$48\%$ excess of the mean over the median is a direct measure of the
skewness discussed in Sect.~\ref{sec:dens}.

Table~\ref{tab:khd} lists the structure vector as a function of the
high-density threshold. The threshold is swept to check that the
working point is not a special choice: we need a value of $k_{\rm HD}$
low enough that the high-density set is populated well enough for its
connected extent to be meaningful, yet high enough that the flagged
galaxies are a genuine overdense population rather than the bulk of the
field. At $k_{\rm HD}=1$, our working value, $500$ of $1963$ galaxies
are flagged high-density and
\begin{equation}
   \Theta = (-0.2112,\; 0.2547,\; 0.1480) .
   \label{eq:theta-work}
\end{equation}
Read component by component, this says that the area-weighted density
sits about $21\%$ below the median (the fixed offset explained next),
that a quarter of the galaxies exceed the high-density threshold, and
that the largest connected high-density structure holds only about
$15\%$ of that flagged population: an overdense set that is present in
quantity but broken into many pieces. That last number is what the
partial-coverage analysis is built to track. We tabulate rather than
plot the sweep because the reader needs the exact working-point values
that seed the rest of the paper, one of the four columns is constant
by construction (see below), and eight threshold rows do not form a
curve worth a panel.

Two features of the sweep require comment. First, $\Delta_\Omega$ is
negative and identical in every row. This is not a transcription
error but a property of the definition: $\Delta_\Omega$ is evaluated
over the whole interior set and never enters the threshold, so on the
full field it takes one value for every $k_{\rm HD}$. Its sign follows
from the convention of Eq.~(\ref{eq:bkg}): because $1/A$ is
right-skewed, the area-weighted density $\rho_\Omega$ falls below the
median density, so a median-referenced $\Delta_\Omega$ is negative for
any field with this skewness. It is a fixed functional of the field,
not an astrophysical claim about underdensity, and what we study is
how well it is recovered from subsets, not its absolute value.

Second, $L_{\rm HD}$ is small at every threshold, between $0.06$ and
$0.20$: the high-density set is fragmented rather than forming one
dominant structure, and it does not become more connected as the
threshold is raised. The non-monotonicity at large $k_{\rm HD}$ (the
uptick to $0.111$ at $k_{\rm HD}=10$) reflects small-number statistics
on the $27$ surviving galaxies. A patch of this size accordingly
supports thresholds up to $k_{\rm HD}\simeq3$ before the morphological
component becomes noise-dominated.

\begin{table}
\caption{The overdensity structure vector as a function of the high-density
threshold, on the $N=2000$ patch with $1963$ interior carriers.}
\label{tab:khd}
\centering
\begin{tabular}{rrrrr}
\toprule
$k_{\rm HD}$ & $n_{\rm HD}$ & $f_{\rm HD}$ & $L_{\rm HD}$ &
$\Delta_\Omega$ \\
\midrule
 0.0 & 981 & 0.4997 & 0.1978 & $-0.2112$ \\
 0.5 & 687 & 0.3500 & 0.1368 & $-0.2112$ \\
 1.0 & 500 & 0.2547 & 0.1480 & $-0.2112$ \\
 1.5 & 359 & 0.1829 & 0.0836 & $-0.2112$ \\
 2.0 & 272 & 0.1386 & 0.0919 & $-0.2112$ \\
 3.0 & 175 & 0.0891 & 0.0743 & $-0.2112$ \\
 5.0 &  90 & 0.0458 & 0.0556 & $-0.2112$ \\
10.0 &  27 & 0.0138 & 0.1111 & $-0.2112$ \\
\bottomrule
\end{tabular}
\end{table}

\subsection{Information distance versus coverage}
\label{sec:res-dg}

The pooled deviation scales entering Eq.~(\ref{eq:g}) are
$(0.0249,\,0.0182,\,0.0412)$ for $(\Delta_\Omega, f_{\rm HD},
L_{\rm HD})$: the morphological component is intrinsically the most
variable under subsetting, by a factor of about two relative to
$f_{\rm HD}$, before any normalisation.

\begin{figure*}
   \centering
   \includegraphics[width=\hsize]{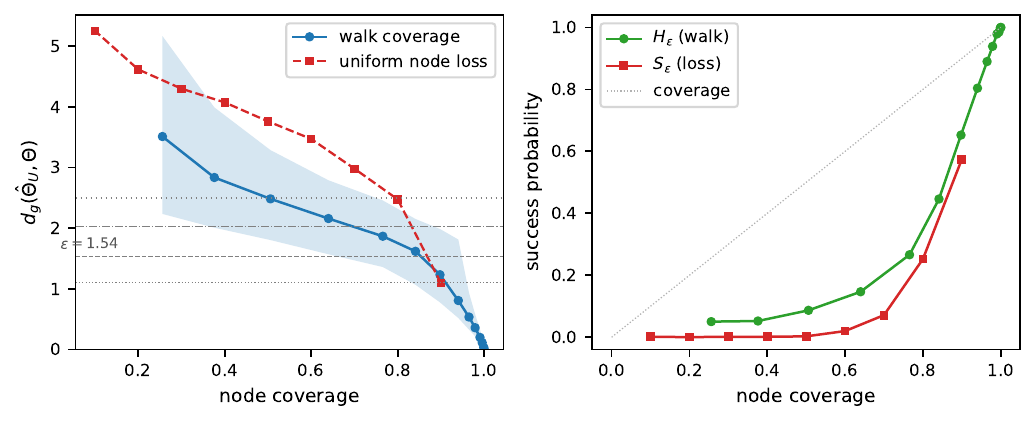}
   \caption{\emph{Left}: median information distance $d_g$ between the
   subset estimate and the full-field structure vector, as a function of node
   coverage, for walk coverage (circles, with $16$--$84$ percentile
   band) and uniform node loss (squares). Grey horizontal lines mark
   the $\varepsilon$ family of Sect.~\ref{sec:success}. \emph{Right}:
   the resulting inferability $H_\varepsilon$ and survivability
   $S_\varepsilon$ at $\varepsilon=1.54$, using the Jeffreys-smoothed
   estimator. The dotted diagonal is the coverage fraction itself,
   shown for reference. The two curves are not related by
   complementation. At the primary tolerance the walk mechanism has
   the larger joint success probability, but the ordering is tolerance
   dependent and reverses at $\varepsilon=1.10$
   (Sect.~\ref{sec:res-mech}); the metric-independent mechanism
   contrast is component-wise and is shown in Fig.~\ref{fig:rmse}.}
   \label{fig:curves}
\end{figure*}

The left panel of Fig.~\ref{fig:curves} shows the median information
distance as a function of coverage for both mechanisms, with the
$16$--$84$ percentile band for walk coverage. Distances fall
monotonically to zero, and at complete coverage $d_g$ vanishes
identically, confirming the consistency requirement of
Sect.~\ref{sec:success}. The two mechanisms separate at intermediate
coverage: the walk median
is $2.16$ at $64\%$ coverage and $1.87$ at $77\%$, against $2.98$ for
uniform loss at $70\%$ and $2.48$ at $80\%$. Uniform loss at $80\%$
coverage is thus more damaging in information terms than walk coverage
at $64\%$. The ordering does not persist to high coverage: by
$90\%$ the loss median has fallen to $1.10$, below the walk value of
$1.23$, and both must vanish at completeness. We return to this
crossing in Sect.~\ref{sec:res-mech}.

\subsection{Inferability and survivability}
\label{sec:res-HS}

Table~\ref{tab:H} gives the inferability at the four tolerance values.
The right panel of Fig.~\ref{fig:curves} already plots $H$ at the
primary tolerance; we tabulate the full family because the tolerance
dependence, and in particular the crossing with survivability
discussed in Sect.~\ref{sec:res-mech}, turns on the exact entries at
each $\varepsilon$, which four overlaid curves per mechanism would
obscure rather than convey. The median distance in the third column
serves the same purpose, fixing the numerical scale behind each row.
At the primary tolerance $\varepsilon=1.54$, $H$ rises from $0.05$ at
$26\%$ coverage through $0.15$ at $64\%$, $0.27$ at $77\%$, $0.45$ at
$84\%$, $0.65$ at $90\%$ and $0.80$ at $94\%$, reaching $0.98$ at
$99\%$. The grid is deliberately dense through this range, so the rise
is measured rather than interpolated: some four fifths of the total
gain occurs between $\simeq65\%$ and $\simeq95\%$ coverage. Below that
range the vector is not recoverable at any tolerance we consider, and
above it recovery is near-certain.

The shape of the tolerance family is informative in its own right. At
$26\%$ coverage, relaxing the tolerance from $\varepsilon=1.10$ to
$2.50$ raises the success probability from $0.02$ to $0.22$, whereas
at $90\%$ coverage the same relaxation moves it from $0.41$ to
$0.97$.
A tolerance choice therefore buys much more at high coverage than at
low, because at low coverage the deviations are not merely large but
broadly distributed.

\begin{table}
\caption{Inferability $H_\varepsilon$ under walk coverage, at the four
tolerances of Sect.~\ref{sec:success}, with the median information
distance. $W$ is the number of walkers of length $T=16$; $M=2000$
realisations per row.}
\tablefoot{Probabilities are Jeffreys-smoothed, $(k+\tfrac12)/(M+1)$,
so for $M=2000$ they lie in $[0.00025, 0.99975]$; entries displayed as
$0.000$ or $1.000$ are rounded and are not exact boundary values.}
\label{tab:H}
\centering
\begin{tabular}{rrrrrrr}
\toprule
$W$ & cover & $d_g^{\rm med}$ & $H_{1.10}$ & $H_{1.54}$ & $H_{2.03}$
& $H_{2.50}$ \\
\midrule
50 & 0.256 & 3.512 & 0.017 & 0.050 & 0.122 & 0.220 \\
80 & 0.376 & 2.837 & 0.013 & 0.052 & 0.166 & 0.350 \\
120 & 0.506 & 2.484 & 0.019 & 0.087 & 0.254 & 0.513 \\
175 & 0.640 & 2.160 & 0.035 & 0.147 & 0.413 & 0.706 \\
250 & 0.766 & 1.868 & 0.073 & 0.266 & 0.620 & 0.849 \\
320 & 0.841 & 1.620 & 0.182 & 0.446 & 0.782 & 0.944 \\
400 & 0.898 & 1.234 & 0.411 & 0.652 & 0.858 & 0.974 \\
500 & 0.940 & 0.809 & 0.728 & 0.804 & 0.894 & 0.987 \\
600 & 0.965 & 0.537 & 0.873 & 0.890 & 0.920 & 0.989 \\
700 & 0.979 & 0.361 & 0.935 & 0.939 & 0.947 & 0.995 \\
850 & 0.990 & 0.200 & 0.976 & 0.979 & 0.980 & 0.998 \\
1000 & 0.995 & 0.116 & 0.983 & 0.985 & 0.985 & 0.997 \\
1300 & 0.999 & 0.039 & 0.997 & 0.998 & 0.998 & 0.999 \\
2000 & 1.000 & 0.000 & 1.000 & 1.000 & 1.000 & 1.000 \\
4000 & 1.000 & 0.000 & 1.000 & 1.000 & 1.000 & 1.000 \\
\bottomrule
\end{tabular}
\end{table}

Survivability under uniform loss (Table~\ref{tab:S}) is lower at the
primary tolerance. Retaining $90\%$ of the field succeeds $57\%$ of
the time, $80\%$ retention succeeds $25\%$ of the time, and by $60\%$
retention the vector is effectively unrecoverable ($S=0.02$). The
comparison with $H$ at matched coverage is deferred to
Sect.~\ref{sec:res-mech}, because it depends on the tolerance in a way
that Table~\ref{tab:S} makes visible: at $\varepsilon=1.10$ the $90\%$
entry is $0.50$, above the corresponding walk value of $0.41$.

\begin{table}
\caption{Survivability $S_\varepsilon$ under independent node loss at
rate $p$, at the same four tolerances as Table~\ref{tab:H}, with the
median information distance. $M=2000$ realisations per row.}
\label{tab:S}
\centering
\begin{tabular}{rrrrrrr}
\toprule
$p$ & cover & $d_g^{\rm med}$ & $S_{1.10}$ & $S_{1.54}$ & $S_{2.03}$
& $S_{2.50}$ \\
\midrule
0.1 & 0.900 & 1.099 & 0.501 & 0.574 & 0.624 & 0.754 \\
0.2 & 0.800 & 2.476 & 0.173 & 0.252 & 0.332 & 0.510 \\
0.3 & 0.700 & 2.982 & 0.037 & 0.071 & 0.130 & 0.263 \\
0.4 & 0.600 & 3.473 & 0.007 & 0.020 & 0.049 & 0.117 \\
0.5 & 0.500 & 3.762 & 0.001 & 0.003 & 0.015 & 0.052 \\
0.6 & 0.400 & 4.074 & 0.001 & 0.001 & 0.005 & 0.023 \\
0.7 & 0.300 & 4.300 & 0.000 & 0.001 & 0.005 & 0.017 \\
0.8 & 0.200 & 4.617 & 0.000 & 0.000 & 0.002 & 0.019 \\
0.9 & 0.100 & 5.253 & 0.000 & 0.001 & 0.008 & 0.033 \\
\bottomrule
\end{tabular}
\tablefoot{Comparison with Table~\ref{tab:H} at matched coverage is
the subject of Sect.~\ref{sec:res-mech}; note that the ordering of $H$
and $S$ reverses between $\varepsilon=1.10$ and $\varepsilon=1.54$.
Per-component relative errors are shown in Fig.~\ref{fig:rmse}.}
\end{table}

\subsection{The component hierarchy}
\label{sec:res-hier}

\begin{figure*}
   \centering
   \includegraphics[width=\hsize]{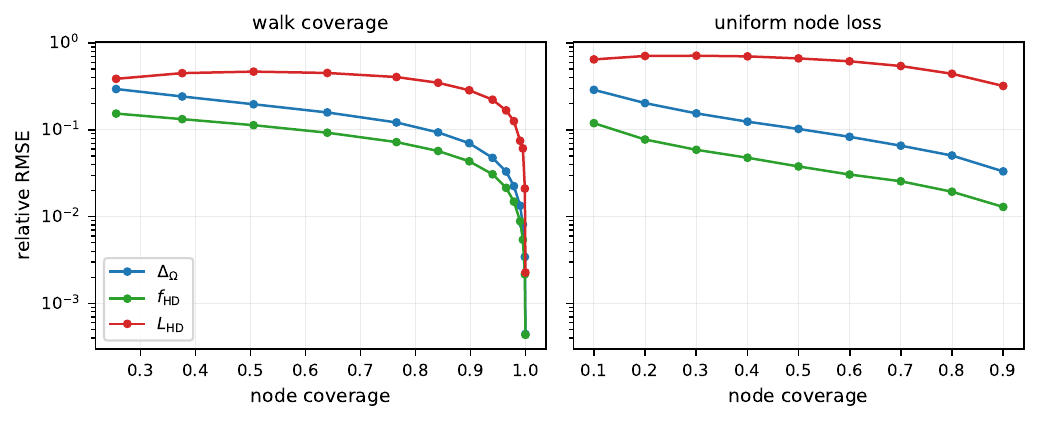}
   \caption{Relative RMSE $\delta_a$ of each component of the structure
   vector (Eq.~\ref{eq:delta-a}) against node
   coverage, for walk coverage (left) and uniform
   node loss (right). The ordering is invariant: the connected extent
   $L_{\rm HD}$ is recovered several times less accurately than the
   scalar components at every incomplete coverage sampled and under both
   mechanisms, by a factor that widens as coverage improves.
   Note the opposite mechanism preference of the two families of
   curves, discussed in Sect.~\ref{sec:res-mech}. The point at
   complete coverage is omitted from the left panel: the RMSE
   vanishes identically there and cannot be displayed on a
   logarithmic axis.}
   \label{fig:rmse}
\end{figure*}

Figure~\ref{fig:rmse} decomposes the information distance into the
relative accuracy of each component. For component $a$ we measure the
relative root-mean-square error
\begin{equation}
   \delta_a = \frac{\sqrt{\mathbb{E}\bigl[(\hat\Theta_a-\Theta_a)^2
   \bigr]}}{|\Theta_a|} ,
   \label{eq:delta-a}
\end{equation}
where the absolute value in the denominator is needed because
$\Delta_\Omega<0$. The ordering is the same
throughout the sampled incomplete-coverage range and for both
mechanisms: $f_{\rm HD}$ is recovered most
accurately, $\Delta_\Omega$ next, and $L_{\rm HD}$ least, by a wide
margin. Under walk coverage at $64\%$, the relative RMSE values are
$0.093$, $0.158$ and $0.451$ respectively; at $90\%$ they are $0.043$,
$0.070$ and $0.285$; at $99.5\%$ they are $0.005$, $0.008$ and
$0.061$. The gap does not merely persist but widens as coverage
improves: measured against $f_{\rm HD}$, the morphological component
is a factor $2.5$ worse at a quarter coverage, $6.6$ worse at $90\%$
and about $10$ worse at $99.5\%$. It remains at the several-per-cent
level when the scalars have converged to a fraction of a per cent, and
is in that precise sense the last component to converge.

This is the expected consequence of the progression built into
$\Theta$ (Sect.~\ref{sec:theta}), but the size of the effect is worth
emphasising. An observer reporting a high-density fraction from a
$64\%$-covered field is making a $9\%$ error; an observer reporting
the connected extent of the same high-density set from the same field
is making a $45\%$ error. These are not comparable measurements, and
quoting a common coverage fraction for both conceals the difference
entirely.

\subsection{Mechanism dependence}
\label{sec:res-mech}

The comparison at matched coverage is the central result, and it must
be read at the component level rather than through a single
probability. At coverage $0.90$ --- $W=400$ walkers giving $0.898$,
against $p=0.1$ loss giving $0.900$ --- the joint criterion is met
with probability $H=0.65$ for the walk subsets and $S=0.57$ for the
uniform subsets, a factor $1.1$ at identical subset size.

That ordering is not uniform in the tolerance, and the reason is
instructive. Uniform loss has the \emph{smaller} median distance at
this coverage ($1.10$ against $1.23$) but a heavier tail, while walk
coverage is typically further away yet more tightly concentrated.
Comparing Tables~\ref{tab:H} and~\ref{tab:S} at matched coverage, the
ordering therefore reverses within our own tolerance family: at
$\varepsilon=1.10$ the uniform subsets succeed more often
($0.50$ against $0.41$), while for $\varepsilon \ge 1.54$ the walk
subsets do. The joint criterion is a statement about the dispersion of
two error distributions at a stated tolerance, not a claim that walk
subsets are generally closer to the full-field structure vector. It is also
sensitive to the calibration of $g_\Theta$, which is generated by the
walk mechanism (Sect.~\ref{sec:metric}); under a diagonal
approximation, which discards the correlation structure inherited from
that ensemble, the ordering reverses as well.

The component-wise comparison below carries the mechanism-level
result, and it is independent of $g_\Theta$, of $\varepsilon$ and of
the calibration ensemble. What survives is not that one mechanism is
uniformly better, but that the two are biased in opposite directions:
inferability is not a function of coverage alone.

The per-component decomposition shows the two mechanisms are not
simply better and worse, but differently biased. At the same matched
coverage of $0.90$, the relative RMSE values are those of
Table~\ref{tab:mech}.

\begin{table}
\caption{Per-component relative RMSE $\delta_a$ (Eq.~\ref{eq:delta-a})
at matched coverage $0.90$, for the two subset mechanisms. Uniform
loss is more accurate on the two scalar components, walk coverage on
the morphological component $L_{\rm HD}$: the mechanisms are biased in
opposite directions, not uniformly better or worse.}
\label{tab:mech}
\centering
\begin{tabular}{lccc}
\toprule
mechanism & $\delta\Delta_\Omega$ & $\delta f_{\rm HD}$ &
$\delta L_{\rm HD}$ \\
\midrule
walk coverage      & 0.070 & 0.043 & 0.285 \\
uniform node loss  & 0.033 & 0.013 & 0.320 \\
\bottomrule
\end{tabular}
\end{table}

Uniform loss recovers the scalar components about twice as accurately
($f_{\rm HD}$ more than three times), while walk coverage recovers the
morphology better. The same reversal holds lower down, where loss at
$70\%$ coverage attains $\delta\Delta_\Omega=0.066$ against $0.158$
for the walk at $64\%$ but $\delta L_{\rm HD}=0.542$ against the walk's
$0.451$, and again at $50\%$; it is therefore a property of the
mechanisms rather than of a particular operating point.

The explanation is geometric. Independent loss assigns the same
inclusion probability to every carrier and therefore avoids the
degree- and density-dependent selection bias of walk coverage. Its
scalar estimates consequently carry much smaller systematic error,
although finite-sample variance remains and $\Delta_\Omega$, being a
ratio involving the inverse of a sample mean, is not exactly unbiased
even under this mechanism. But uniform
removal punches holes throughout the field, and a high-density
component survives as one connected object only if none of the
galaxies bridging it is removed; connectivity is therefore attacked
everywhere at once. Walk coverage does the opposite. It is a biased sample,
over-representing high-degree and therefore locally dense galaxies,
which is why its scalar estimates are worse. But short walks
preferentially preserve local adjacency chains within the regions they
visit while leaving spatially clustered gaps elsewhere: at $90\%$
coverage the unvisited complement breaks into components of which the
largest holds about $9\%$ of the missing galaxies, against about $3\%$
under uniform loss.

The weighting in the joint criterion works in the opposite direction
to the one this might suggest. In a Mahalanobis form a component with
larger reference scale receives a \emph{smaller} precision weight, so
$L_{\rm HD}$, having the largest pooled deviation scale, is the most
strongly downweighted of the three. Its raw error is nevertheless
large enough in proportion that the standardised term still dominates
the budget under both mechanisms: at matched coverage the ratios of
root-mean-square deviation to reference scale are $1.06$ for walk and
$1.21$ for uniform loss, against below $0.6$ for either scalar under
either mechanism. The off-diagonal structure of $g_\Theta$ then acts
asymmetrically,
inflating walk distances by about $15\%$ relative to a diagonal
approximation but loss distances by about $50\%$, because the
reference correlations are inherited from the walk ensemble. Both
effects are why $H$ and $S$ cannot be complementary: they are
probabilities under mechanisms that stress different parts of the
structure vector and that the reference metric does not treat
symmetrically.

\subsection{Categorical geometry}
\label{sec:res-cat}

Evaluating Eq.~(\ref{eq:cat}) at the matched-coverage operating point
$(H,S)=(0.65,0.57)$ with $\alpha=1/2$ gives $g_{\alpha\alpha}=4.00$,
$g_{SS}=2.05$ and $g_{HH}=2.20$. Both structural components are of
comparable magnitude and both are close to the minimum of the
Bernoulli factor $1/[p(1-p)]$, which is attained at $p=1/2$. The
model is therefore locally least sensitive here: neighbouring values
of $H$ or $S$ are the hardest to distinguish in Fisher distance, so a
small change in either is the most difficult to detect. This is the
regime in which the trade-off between the two is real rather than
nominal, and it is where a figure of merit such as Eq.~(\ref{eq:fsec}) or
Eq.~(\ref{eq:fastro}) carries the most information about a design
choice.

At operating points nearer the boundaries the metric behaves very
differently. At $50\%$ retention, where $S=0.003$, the same evaluation
gives $g_{SS}=182$: configurations with nearly identical survivability
are sharply distinguishable there in Fisher distance. That
discriminating power is of little practical use, however, since every
configuration in that regime fails the criterion almost surely. The
divergence is the reason the Jeffreys smoothing of Sect.~\ref{sec:HS} is
required and not
cosmetic: an exact zero would make the metric singular rather than
merely large.

\subsection{The inverse-probability weight fails consistency}
\label{sec:res-ipw}

We ran the full analysis with degree-based inverse-probability
weights, $w_i = 1/\pi_i \propto 1/\deg(i)$, in
Eqs.~(\ref{eq:delta})--(\ref{eq:lhd}). The result fails the
consistency check of Sect.~\ref{sec:success}: at complete coverage,
where the unweighted estimator gives $d_g=0$ exactly, the weighted
estimator gives $d_g = 3.79$, with a residual relative bias of
$0.28$ in $\Delta_\Omega$ and $0.17$ in $f_{\rm HD}$ that does not
decrease as coverage grows. The weighting itself is mild: on the full
interior set the degree
weights retain an effective sample size $N_{\rm eff}=1856$ out of
$n=1963$, or $95\%$. The failure is therefore one of what is being
estimated rather than of variance inflation. The
weighted curve is
flat in this
respect: the residual is $3.76$ at $99.9\%$ coverage and $3.79$ at
both $100\%$ points, so the estimator converges to the wrong value
rather than converging slowly.

This is the behaviour predicted in Sect.~\ref{sec:ipw}. The
degree weight corrects a frequency-valued sample, and our observable
is set-valued; the mismatch is a genuine error in what is being
estimated rather than a tuning problem, and no choice of $\lambda$ or
$\varepsilon$ repairs
it. The diagnostic value of the consistency identity is worth noting
here: had we examined only the low-coverage end of the curve, where
the weighted estimator gives plausible-looking numbers, the error
would not have been visible. We therefore report unweighted estimates
throughout and defer the inclusion-probability weight, which satisfies
$\pi^{\rm incl}_i \to 1$ and hence the consistency identity by
construction, to future work.

   \section{Discussion}
   \label{sec:discussion}

The results support a single structural claim: partial coverage
degrades a structure measurement anisotropically in information space,
and the direction of the degradation is set by the sampling mechanism
rather than by its severity.

The immediate practical consequence is that coverage or completeness
alone cannot summarise what a partially observed field supports. Two
subsets of identical size, drawn by mechanisms that are
both entirely reasonable models of real incompleteness, differ in
opposite directions component by component: one preserves amplitudes
and destroys connectivity, the other does the reverse. The two
mechanisms should be read as idealised limiting cases rather
than as models of any particular survey. Independent node loss
represents random object-level incompleteness
(catalogue thinning, detection failure, spatially uncorrelated
selection), and a survey dominated by it will report accurate
amplitudes and unreliable
connectivity. Bright-star masks, chip gaps and unfinished tilings are
instead spatially correlated mechanisms, closer in character to walk
coverage, and a survey dominated by them will report the opposite.
Real incompleteness is usually a mixture, and its behaviour may lie
between the two limits studied here. Both may quote the same
completeness.

The component hierarchy also has a bearing on how structure statistics
are chosen. Connectivity-based statistics --- filament identification,
percolation analyses, friends-of-friends membership, void topology ---
are all of the same character as $L_{\rm HD}$, in that they are
defined through adjacency of surviving members rather than through
averages over them. Our measurements suggest such statistics carry
several times the partial-coverage error of amplitude statistics
computed from the same catalogue, which argues for reporting them with
mechanism-specific rather than generic incompleteness corrections.

Three limitations bound these conclusions. First, the analysis is
two-dimensional and run on a thin redshift slice; the balance between
mechanisms could shift in three dimensions, where connectivity has
more routes available and may be correspondingly more robust to
uniform removal. Second, the background of Eq.~(\ref{eq:bkg}) is held
fixed across subsets, which isolates the recovery question but does not
address calibration stability. An observer who must estimate
the background from the same partial field faces an additional and
possibly larger error. Third, the patch is $0\fdg23$ on a side and
supports thresholds only up to $k_{\rm HD}\simeq3$
(Sect.~\ref{sec:res-field}); the sharpest overdensities of the full
field are not represented in the morphological component at this
patch size.

Two natural extensions follow from these limitations and results. The
first is the correct sampling
correction. Section~\ref{sec:res-ipw} establishes that the degree weight is
inconsistent for set-valued coverage and identifies the inclusion
probability $\pi^{\rm incl}_i = \Pr[i \in U]$ as the right first-order
design object \emph{for the scalar target quantities}; it is estimable
from
the same Monte Carlo ensemble that produces the curves, at no
additional sampling cost, and the consistency identity
$d_g(V^{\rm int})=0$ provides a sharp test of whether the resulting
estimator is correct. Correcting $L_{\rm HD}$ would additionally
require joint inclusion information or an explicit model of the
missing connectivity.

The second couples this analysis to graph reconstruction. Here the
adjacency is known exactly, and the only source of error is which
galaxies are observed. When the adjacency must instead be inferred
from observed traffic \citep{imbrisak_paperII}, a second error enters,
and the two are not independent: walk coverage controls both which
galaxies are seen and which edges the reconstruction can recover.
Running the present analysis on a reconstructed adjacency $\hat{G}$,
with subset realisations matched between $G$ and $\hat{G}$ so that the
comparison is paired, would separate the coverage-limited from the
reconstruction-limited contributions to $H_\varepsilon$. The
reconstruction is computationally the binding constraint, since
its per-iteration cost is dominated by a finite-difference Jacobian
and a pseudoinverse. That experiment therefore belongs on a smaller
patch nested inside the present one, which is why the two analyses are reported
separately. The question it would answer is a specific one: whether the
overdensity structure vector stabilises before the graph
is perfectly reconstructed, as the coverage-limited behaviour of
\citet{imbrisak_paperII} suggests it should.

One caveat travels with the framework rather than with this dataset.
A joint criterion compresses a vector of errors into one probability,
and the compression is done by a metric that must be calibrated on
something. Our calibration ensemble is generated by one of the two
mechanisms under comparison, and Sect.~\ref{sec:res-mech} shows the
resulting single-number comparison is sensitive both to that choice
and to the tolerance. A symmetric calibration, drawing equally on
every mechanism to be compared, would be the natural remedy where a
scalar summary is wanted; the component-level statement needs no such
choice, which is why we rest the conclusions on it.

Finally, the framework is not specific to the vector we chose. Any
vector of structure statistics can be substituted for $\Theta$, and the
machinery (the pooled-deviation metric, the tolerance family,
the matched-coverage comparison, the consistency identity) carries
over unchanged. The choice of $\Theta$ is where the astrophysics
enters; the rest is a statement about inference from subsets.

   \section{Conclusions}
   \label{sec:conclusions}

We have formulated the recovery of galaxy overdensity structure from a
partially observed field as an inference problem on a fixed
structure vector, evaluated on a $z$-sliced COSMOS field. The wider
lesson is that how much of a field was observed and how much of its
structure was preserved are different quantities, and only the second
governs what the field can support. Our conclusions are as follows.

\begin{enumerate}
   \item Structure statistics do not degrade uniformly under partial
   coverage. Across the vector we study, morphology --- the connected
   extent of the high-density set --- is systematically the hardest
   component to recover and the last to converge, several times less
   accurate than the amplitude components at matched coverage, with the
   gap widening as coverage improves. Statistics defined through the
   adjacency of surviving members are intrinsically more fragile than
   those defined through averages over them.

   \item Inferability is not a function of coverage alone but also of
   how the field is sampled. Two mechanisms of the same severity bias
   the measurement in opposite directions --- one preserves amplitudes
   and erodes connectivity, the other the reverse --- so a single
   coverage fraction cannot summarise what a partial field supports.
   This mechanism dependence is stated at the component level, where it
   is independent of any metric; a joint single-number criterion can
   order the mechanisms but does so in a tolerance-dependent way and is
   the weaker statement.

   \item Sampling corrections must match the kind of observable. The
   degree-based inverse-probability weight, natural for frequency-valued
   random-walk sampling, is inconsistent for a set-valued observable and
   leaves a bias that does not vanish at complete coverage. The exact
   recovery identity at full coverage provides a sharp and inexpensive
   test of whether a proposed correction is consistent, and connectivity
   components need joint inclusion information that no per-galaxy weight
   supplies.

   \item The framework is not specific to the vector or the dataset
   chosen. Any vector of structure statistics can be substituted, and
   the same machinery --- the pooled-deviation reference metric, the
   tolerance family, the matched-coverage comparison and the
   consistency identity --- carries over. The choice of structure vector
   is where the astrophysics enters; the rest is a statement about
   inference from subsets.
\end{enumerate}

\begin{appendix}
\nolinenumbers
\section{Semantics of the figure of merit: a security demonstrator}
\label{app:sec}

The pair $(H_\varepsilon, S_\varepsilon)$ and its geometry,
Eq.~(\ref{eq:cat}), are indifferent to whether partial access is
something one wants. The astrophysical reading of
Sect.~\ref{sec:results} treats inferability as a benefit; the
distributed-storage reading of \citet{zlatic2026} treats the same
quantity as an exposure. Equations~(\ref{eq:fsec})
and~(\ref{eq:fastro}) differ only in that sign. This appendix checks
that the machinery of Sect.~\ref{sec:method} transfers without
modification when the sign is flipped, and it is included as a
demonstration of semantics rather than as a second case study.

We use Zachary's karate club \citep{zachary1977}, a $34$-vertex,
$78$-edge social network with a documented split into two factions of
$17$ members each. Writing $A$ for one faction, we define a security
structure vector in exact analogy with
Eq.~(\ref{eq:product}),
\begin{equation}
   \Theta_{\rm sec}(U) = \bigl(\pi_A(U),\, Q_A(U),\, L_A(U)\bigr),
   \label{eq:thetasec}
\end{equation}
where $\pi_A(U) = |U \cap A|/|U|$ is the observed share of the target
faction, $Q_A(U)$ is the modularity of the known two-way partition evaluated on
the observed subgraph $G[U]$ (a standard measure of how well a
partition separates a network into communities), and $L_A(U)$ is the
share
of the observed faction members carried by their largest connected
component in $G[U \cap A]$. The three components progress from
composition, through mesoscopic
community structure, to within-faction connectivity, mirroring the
progression built into Eq.~(\ref{eq:product}). We set $Q_A(U)=0$ when
$G[U]$ carries no observed edge, and likewise when one of the two
factions is entirely unobserved, since the modularity of the partition
is undefined in both cases. The information restriction of
Sect.~\ref{sec:contract} is unchanged: the estimator sees the vertices
in $U$, the adjacency of $G$ restricted to $U$, and the faction labels
of observed vertices only.

Everything else is taken from Sect.~\ref{sec:method} without change of
procedure: the same pooled-reference-metric construction, re-estimated
from the karate-club deviation ensemble; the same tolerance family
from $\chi^2_3$ quantiles; the same Jeffreys smoothing; and the same
two mechanisms, with walks of length $T=6$ on this much smaller graph
and $M=2000$ realisations per grid point. Walk coverage is swept over
$W \in \{1,2,3,4,6,8,12,20,40\}$ and node loss over
$p \in \{0.1,\dots,0.9\}$ in steps of $0.1$; the calibration ensemble
is $60$ realisations at each of the nine $W$ levels, that is $540$
deviations in total. The reference metric is therefore a different
matrix from the one used in Sect.~\ref{sec:results}, as it must be for a
different structure vector. The
full-network value is
$\Theta_{\rm sec} = (0.500,\, 0.391,\, 1.000)$ and the pooled
deviation scales are $(0.19,\, 0.17,\, 0.20)$.

\begin{figure}[!ht]
   \centering
   \includegraphics[width=\hsize]{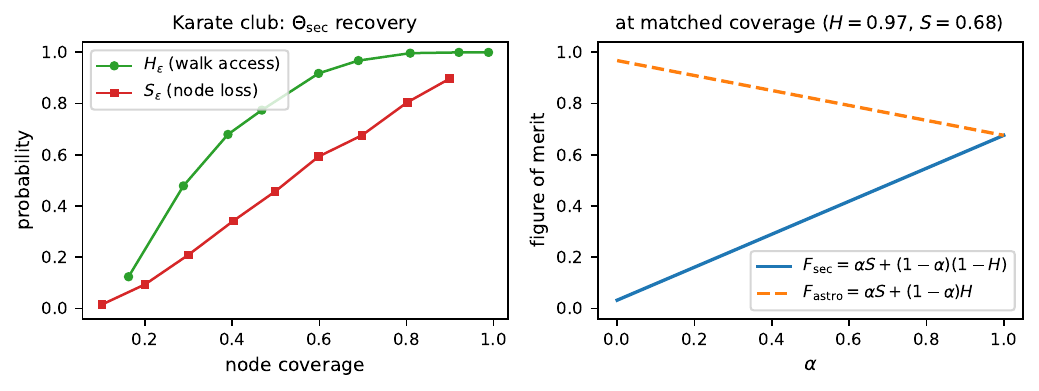}
   \caption{Security demonstrator on the karate club.
   \emph{Left}: inferability under walk access and survivability under
   independent node loss, for the structure vector of Eq.~(\ref{eq:thetasec}) at
   $\varepsilon=1.54$.
   \emph{Right}: the two figures of merit at the matched-coverage
   operating point. The curves are the same object; only the sign
   with which $H$ enters distinguishes them, and the two utilities
   therefore rank the same configuration oppositely at small
   $\alpha$.}
   \label{fig:karate}
\end{figure}

Figure~\ref{fig:karate} shows the result. At a matched coverage of
$0.70$ the walk mechanism recovers $\Theta_{\rm sec}$ with probability
$H=0.97$, while independent loss preserves it with probability $S=0.68$, the
same direction of asymmetry as in the main analysis and for the same
reason, since $L_A$ is a morphological component and
the full-network faction is connected. In the astrophysical reading
that is a favourable configuration; in the security reading it is not,
and the two utilities separate accordingly: at $\alpha=0$,
$F_{\rm sec}=0.03$ against $F_{\rm astro}=0.97$, converging to the
common value $S=0.68$ as $\alpha \to 1$. The categorical metric at
this operating point gives $g_{\alpha\alpha}=4.00$, $g_{SS}=2.28$ and
$g_{HH}=16.0$: the large $g_{HH}$ places the configuration near the
$H \to 1$ boundary, where inferability is sharply determined and there
is correspondingly little left to trade.

Nothing in Sect.~\ref{sec:method} was adapted to obtain this. The
geometry, the estimator architecture, the calibration procedure and
the tolerance convention are inherited from Sect.~\ref{sec:method};
the choice of $\Theta$ is what carries the application, and the choice
of utility is what carries its interpretation.
\end{appendix}


\end{document}